\documentclass[fleqn,usenatbib]{mnras}
\usepackage{newtxtext,newtxmath}
\usepackage{graphicx}
\usepackage{bm}
\usepackage{ulem}

\newcommand{\approptoinn}[2]{\mathrel{\vcenter{
        \offinterlineskip\halign{\hfil$##$\cr
        #1\propto\cr\noalign{\kern2pt}#1\sim\cr\noalign{\kern-2pt}}}}}

\usepackage{color}

\newcommand{\Sec}[1]{\S\ref{#1}}

\newcommand{\Fig}[1]{Figure~\ref{#1}}
\newcommand{\Tab}[1]{Table~\ref{#1}}

\graphicspath{{./fig/}{./png/}}

\title[Evolution of the Sun's magnetic field in Cycles 21--24]
{Evolution of the Sun's activity and the poleward transport of remnant magnetic flux in Cycles 21--24}

\author[Mordvinov et al.]{Alexander  V. Mordvinov$^{1}$, Bidya Binay Karak$^{2}$\thanks{E-mail: karak.phy@iitbhu.ac.in}, Dipankar Banerjee$^{3,4}$, Elena M. Golubeva$^{1}$, \newauthor
Anna I. Khlystova$^{1}$, Anastasiya V. Zhukova$^{5}$, Pawan Kumar$^{2}$\\
$^{1}$Institute of Solar-Terrestrial Physics, Irkutsk, 664033, Russia\\
$^{2}$Department of Physics, Indian Institute of Technology (Banaras Hindu University), Varanasi 221005, India\\
$^{3}$Aryabhatta Research Institute of Observational Sciences, Nainital 263000, Uttarakhand\\
$^{4}$Indian Institute of Astrophysics, Koramangala, Bangalore 560034, India\\
$^{5}$Crimean Astrophysical Observatory, Nauchny 298409, Bakhchisaray, Republic of Crimea
}

\date{Accepted XXX. Received YYY; in original form ZZZ}

\begin{document}
\label{firstpage}
\pagerange{\pageref{firstpage}--\pageref{lastpage}}
\maketitle

\begin{abstract}
Detailed study of the solar magnetic field is crucial to understand
its generation, transport and reversals. The timing of the reversals 
may have implications on space weather
and thus identification of the temporal behavior of the critical surges
that lead to the polar field reversals is important.
We analyze the evolution of solar activity and magnetic flux transport in Cycles 21--24.
We identify critical surges of remnant flux that reach the
Sun's poles and lead to the polar field reversals. We reexamine the
polar field buildup and reversals in their causal relation to the
Sun's low-latitude activity. We further identify the major remnant
flux surges and their sources in the time-latitude aspect. 
We find that special characteristics of individual
11-year cycles are generally determined by the spatiotemporal
organization of emergent magnetic flux and its unusual properties.
We find a complicated restructuring of high-latitude magnetic fields 
in Cycle~21.
The global rearrangements of solar magnetic fields were caused by surges
of trailing and leading polarities that occurred near the activity
maximum. The decay of non-Joy and anti-Hale active regions resulted
in the remnant flux surges that disturbed the usual order in magnetic
flux transport.
We finally show that the leading-polarity surges during cycle minima
sometimes link the following cycle and a collective effect of these
surges may lead to secular changes in the solar activity.
The magnetic field from a Babcock--Leighton dynamo model generally agrees with these observations.
\end{abstract}

\section{Introduction}
\label{sec:int}
The magnetic field on the surface of the Sun is a fundamental
observable to understand the origin of global magnetism and various heliospheric
processes 
\citep{Hoeksema95, wang09, CCJ07, Priy14, HC19, kumar21,KKV21}.
Some basic properties of the solar magnetic field were recognized
by detecting the active regions from the early measurements \citep{Hale19}
and later by detecting the
spatial and temporal evolution of magnetic field using early
magnetograph \citep{Babcock1955}. Based on these early observations, 
\citet{Babcock1961}
and \citet{Leighton1964}
provided an empirical concept for cyclic rearrangement
of magnetic field on the solar surface. The decay of long-lived
active regions (ARs) form the unipolar magnetic regions (UMRs).
Then UMRs of predominantly trailing polarity are transported
poleward by meridional flow, annihilate the old polarity field,
and develop the new polarity field by continuous supply of trailing
polarities from low latitudes. The latitudinal dependence of 
the tilt angles of bipolar ARs, i.e., Joy's law plays a crucial role in this process. 

The decay of
non-Joy and anti-Hale ARs disturbs the usual order of the poleward
flux transport and leads to the polar field weakening 
\citep{Ca13, MNL14, HCM17, LC17, Nagy17, KM17, KM18, Kit18, wang20}.
In many bipolar magnetic
groups, leading-polarity spots appear at higher latitudes than
the trailing-polarity ones and thus they have negative tilts \citep[e.g.,][]{MNL14,
yeates15}. 
The decay of these
groups leads to the formation of leading-polarity UMRs at
higher latitudes. 

As an 11-year cycle progresses, the polar fields reverse at about
activity maximum. In each hemisphere, this usually happens once in
a cycle and sometimes asynchronously in the northern and southern
hemispheres. \citet{Makarov1983} studied the poleward migration of
chromospheric filaments using long-term observations from Kodaikanal
Solar Observatory (KoSO). Analyzing the chromospheric proxy data, they
studied the global evolution of magnetic fields and found triple
polar field reversals in Cycles 16, 19, and 20 in the northern
hemisphere. The physical nature of multiple polar field reversals
is still poorly known.  

Recently, \citet{Mordvinov2020} reconstructed the solar magnetic field 
using the Sun's emission in the CaII~K and H$\alpha$ lines from KoSO. 
This reconstruction enabled us to study the evolution of solar magnetic
field and reversals in Cycles 15--19.  
The time-latitude analyses of synoptic maps from several
observatories also demonstrated the evolution of photospheric
magnetic fields and their long-term changes \citep{Petrie15, PE17, VM19, KMB18, Janardhan18}. 
Recent analysis of low-resolution synoptic maps from Wilcox Solar Observatory (WSO) revealed a complicated polar field reversal in Cycle~21 \citep{Mordvinov2019}.

Taking into account the previous findings, we reexamine high-resolution
synoptic maps to study the remnant magnetic flux, its poleward transport,
and inter-cyclic surges in Cycles~21--24.
These surges are of fundamental
importance because they link adjacent solar cycles in pairs and
causes a long-term memory in solar dynamo. 
Finally, we show a snapshot of the spatiotemporal evolution
of the magnetic field from a recent three-dimensional (3D) Babcock--Leighton 
dynamo model \citep{KM17} which produces
some features of the solar magnetic field, including opposite polarity surges,
in great detail.

\begin{table*}
 \caption{
Analyzed synoptic maps.
}
 \label{tab:data}
 \begin{tabular}{cccccl}
   \hline
   Data source & Spectral line & \multicolumn{2}{c}{Duration} & Map size & References \\
 &   & Years & CRs & & \\
   \hline
   \multicolumn{6}{c}{\textit{Synoptic maps of magnetic field}} \\
   \hline
   NSO/KPVT$^1$  & Fe I 8688 \AA & 1975--2003 & 1625--2007 & 360$\times$180   & \citet{Jones1992}           \\
   SOLIS/VSM$^2$ & Fe I 6302 \AA & 2003--2012 & 2007--2127 & 360$\times$180   & \citet{Keller2003}          \\
       &         &            &            &                  & \citet{Balasubramaniam2011} \\
   SoHO/MDI$^3$  & Ni I 6768 \AA & 1996--2010 & 1909--2104 & 3600$\times$1080 & \citet{Scherrer1995}        \\
   SDO/HMI$^4$   & Fe I 6173 \AA & 2010--2021 & 2096--2240 & 3600$\times$1440 & \citet{Scherrer2012}        \\
   \hline
   \multicolumn{6}{c}{\textit{Synoptic maps of coronal holes}} \\
   \hline
  NSO/KPVT$^5$  & He I 10830 \AA & 1976--1986 & 1636--1783  & 360$\times$180              & \citet{Jones1992}           \\
   \hline
   \multicolumn{6}{l}{$^1$ \url{https://nispdata.nso.edu/ftp/kpvt/synoptic/}} \\
   \multicolumn{6}{l}{$^2$ \url{https://solis.nso.edu/0/vsm/crmaps/}} \\
   \multicolumn{6}{l}{$^3$ \url{http://soi.stanford.edu/magnetic/synoptic/carrot/M_Corr/}} \\
   \multicolumn{6}{l}{$^4$ \url{http://jsoc.stanford.edu/data/hmi/synoptic/}} \\
   \multicolumn{6}{l}{$^5$ \url{http://158.250.29.123:8000/web/CoronalHoles/FITS/}} \\
\end{tabular}
\end{table*}

\section{Data}

We analyze the homogenized series of Carrington synoptic maps from the National Solar Observatory/Kitt Peak Vacuum Telescope (NSO/KPVT) 
and from the Synoptic Optical Long-term Investigations of the Sun/Vector 
Spectro-Magnetograph (SOLIS/VSM). We also analyze synoptic maps of coronal holes (CHs) 
from NSO/KPVT to study the evolution of open magnetic fluxes at the Sun's poles in Cycle 21. 
We investigate the photospheric magnetic flux evolution using high-resolution synoptic maps 
from the Solar and Heliospheric Observatory/Michelson Doppler Imager (SOHO/MDI) 
and from Solar Dynamics Observatory/Helioseismic and Magnetic Imager (SDO/HMI). 
All of the magnetic field maps show distributions of the radial projection of 
measured line-of-sight component. Table~\ref{tab:data} demonstrates main information 
about the maps considered in this study.  
All these maps are homogenized in $360\times180$ pixels for our analyses.
To pay our special attention to polarity reversal at the Sun's poles and for completeness, 
we supplement the analysis of synoptic maps with the consideration of line-of-sight measurements 
of the Sun's polar magnetic fields
with 20~nHz low pass filtered 
 from WSO\footnote{\url{http://wso.stanford.edu/Polar.html}}\citep{Svalgaard1978,Hoeksema1995} available since 1976. 
To identify the possible sources of remnant magnetic fluxes in Cycles~21-24, we used 
the tilt angles from Debrecen Photoheliographic Data catalogue 
of bipolar ARs \citep{BGL2016,GLB2017}. 
To find information about ARs that violate Hale's polarity law (anti-Hale ARs), 
we used a catalogue of Bipolar Magnetic Regions (BMRs) \citep{SW2016} 
and a catalogue of bipolar ARs violating the Hale's polarity law \citep{Zhukova20}. 

\begin{figure*}
\centering
\includegraphics[scale=0.5]{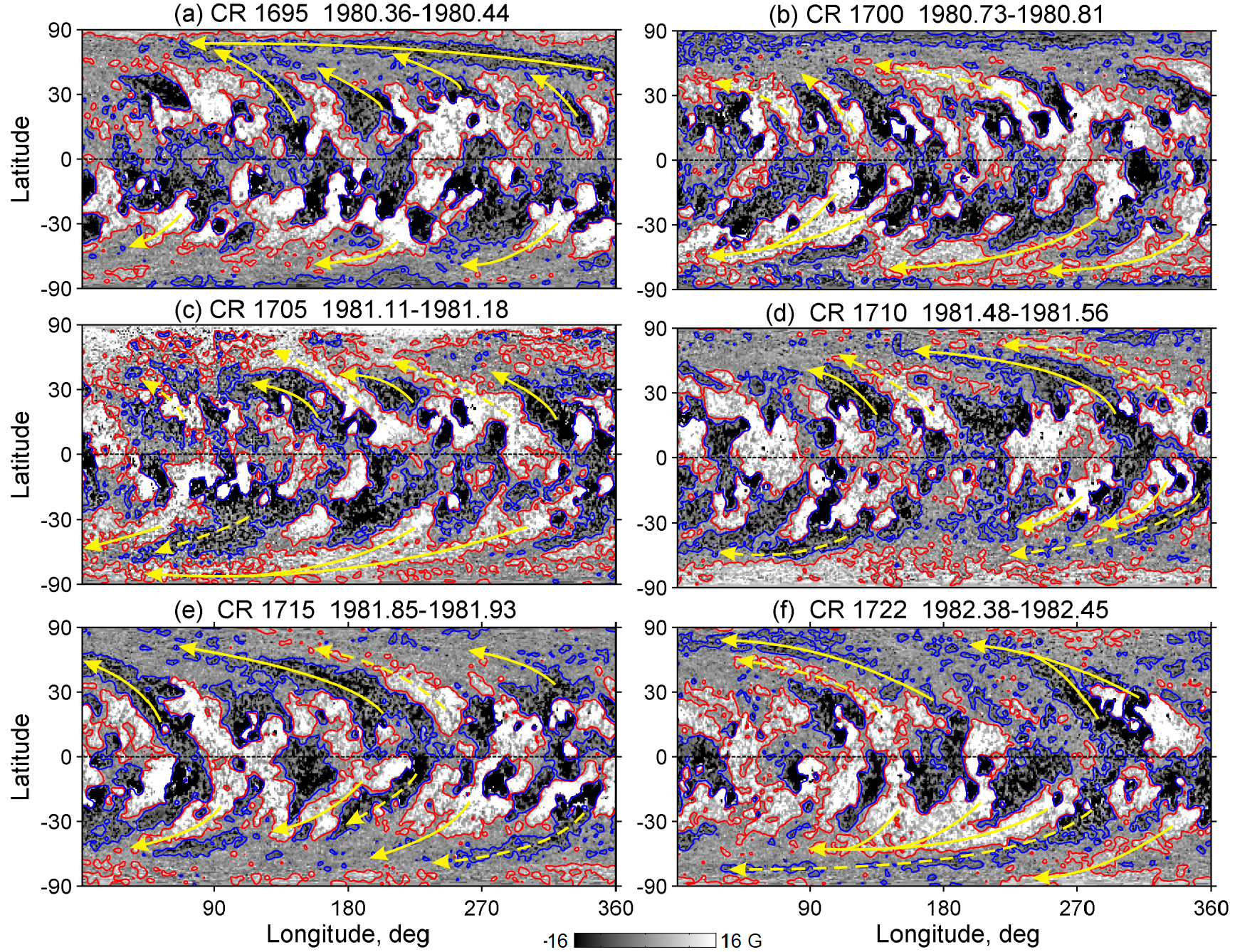}
\caption{
Synoptic maps of magnetic flux for different CRs. 
Positive and negative polarity UMRs are shown in white-to-black. 
The red and blue contours show the boundaries of UMRs corresponding to $\pm 3$~G.
The trailing/leading-polarity surges are marked with solid/dashed arrows. 
}
\label{figure1}
\end{figure*}

\section{Results}
\subsection{A complicated restructuring of solar magnetic field in Cycle 21}
\Fig{figure1} shows original synoptic maps of magnetic flux from NSO/KPVT in black-to-white colors
during the period of polar field reversals in Cycle~21. 
Before the reversal, positive/negative polarities dominated at the north/south poles (\Fig{figure1}a). 
However, at mid-latitudes in the northern hemisphere, 
surges of negative (trailing) polarity were formed. 
At latitudes above $60\degr$, trailing-polarity surges merged in a ring-shaped UMR of negative polarity. 
These surges originated after the decay of ARs.  
Then in Carrington Rotation (CR)~1700, 
northern hemisphere, high-latitude magnetic fields were restructured 
due to their poleward transport (\Fig{figure1}b).
By that time at mid-latitudes, the leading-polarity (positive) surges strengthened 
(dashed arrows). 
These surges formed a tier of the leading-polarity surges,
possibly which resulted in change in the dominant polarity by CR 1705 (\Fig{figure1}c).
We note that the daily magnetograms have regular gaps in the Sun's polar zones due their poor visibility. 
To estimate the missing data, the 
Kitt Peak magnetograms 
and all other data
were filled using the extrapolation technique \citep{Svalgaard78,Sun2011}. 
Such uncertainties sometimes result in serious errors in polar zones \citep{Bertello14}
and thus the data around polar zones should be taken with caution.

The decay of large activity complexes (ACs) in 1981--1982, led to formation of trailing-polarity surges at low- to mid-latitudes. As the cycle progressed, the trailing polarity surges approached the northern polar zone by CR~1710 (solid arrows in \Fig{figure1}d). 
Further strengthening of the trailing-polarity surges and their poleward transport resulted in the third change in dominant polarity at the northern polar zone 
(\Fig{figure1}e,f). 

We note that this peculiar behavior of the polar field in the northern hemisphere and its link with the 
low latitude surges are highlighted by \citet{Ca13} in examples 
of equatorial flux, in particular the situation around 1980.
Using the surface flux transport model, they demonstrated that a single cross-equatorial flux plumes can affect the net hemispheric 
flux of the following minimum by up to $60\%$.  
However, the observational analyses by \citet{PE17} show that large, 
long-lived complexes are the major cause of polar field change.

In the southern hemisphere, surges of trailing- and leading-polarities
were also formed. By CR~1700, the trailing-polarity surges covered
a wide longitude interval. Their further strengthening and the
poleward transport resulted in the polar field reversal at the south
pole by CR 1705. After the decay of several ARs, extensive
leading-polarity surges were formed by CR~1710. During their further
evolution, these surges approached south pole. However, their flux
was weak to change the dominant polarity there.

\begin{figure*}
\centering
\includegraphics[scale=0.42]{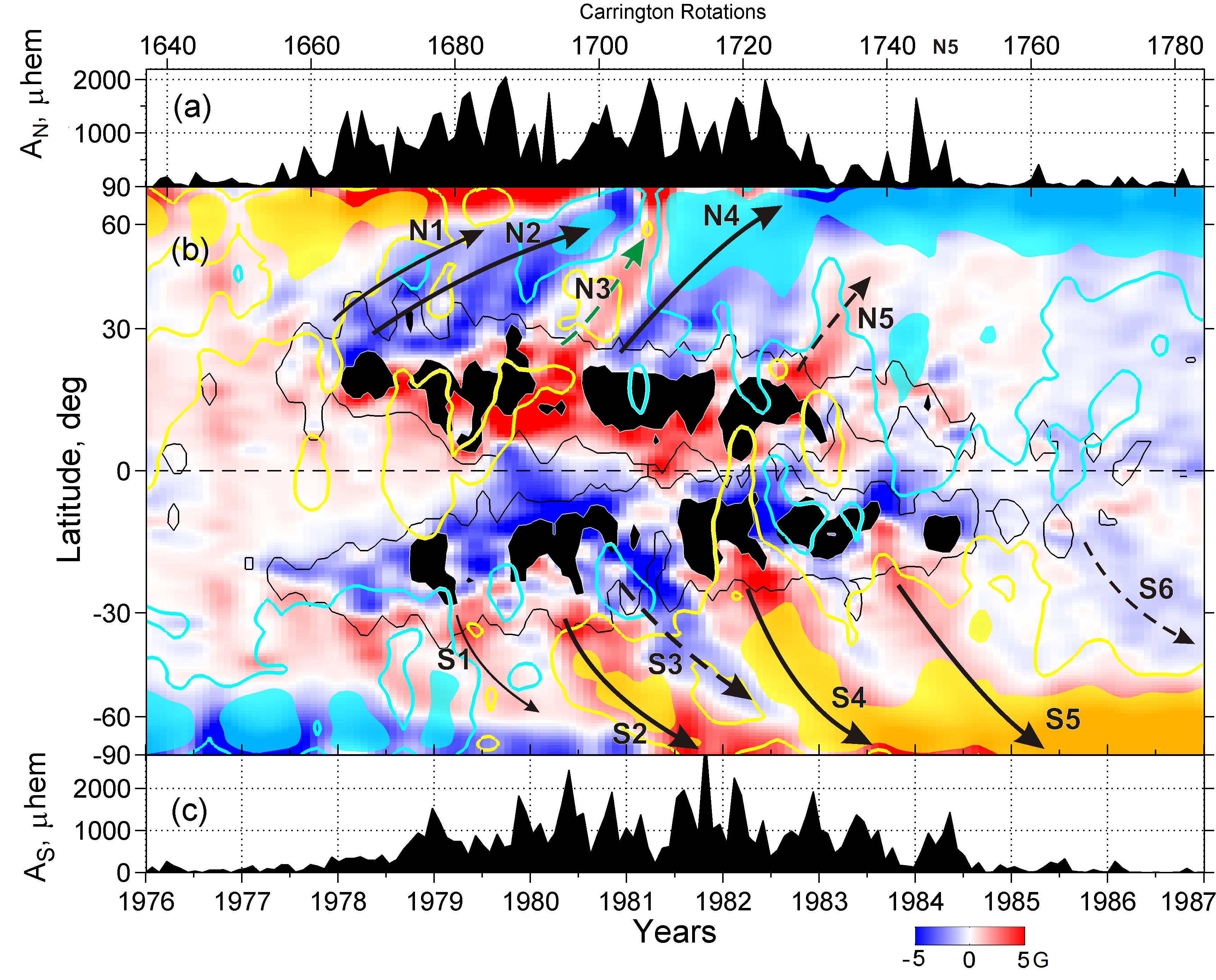}
\caption{
(a/c) Changes in sunspot areas in north/south. 
(b) Time-latitude variation of zonally averaged magnetic fields in the blue-to-red. Domains of high zonal flux density are shown with black spots
and contours. 
Solid/dashed arrows show the trailing/leading-polarity surges. Domains of frequent CH appearances are shown in yellow/cyan color (at levels of  $> 0.1/ < - 0.1$). Yellow/cyan contours correspond to CH appearances at levels $\pm 0.02$. 
}
\label{figure2}
\end{figure*}

The global reorganization of solar magnetic fields 
are completed by
the formation of stable polar coronal holes (PCHs). As the first
remnant-flux of new polarity reaches the polar zones, PCHs of the
preceding cycle disappear. After the polar field reversal, new PCHs
are formed due to the merger of high-latitude  CHs \citep{GM17}.
The polar field buildup in every new cycle occurred in parallel
with new PCH formation. Therefore, the analysis of high-latitude
CHs provides independent information on the
spatiotemporal behavior of polar magnetic fields.


In \Fig{figure2}b, stable polar CHs are shown by yellow/cyan spots within UMRs of positive/negative polarities. 
This distribution demonstrates a macro-structure of the CH ensemble. 
As long-lived ACs evolve and decay, their remnant magnetic fields dissipate and form UMRs. 

It is believed that as the cycle progresses, the following- and leading-polarities UMRs are linked via coronal magnetic arcades \citep{PH13}. At high latitudes, however, magnetic fields tend to open up and locally unbalanced flux patterns arise. High-latitude CHs usually appear within the corridors where the magnetic field opens (\Fig{figure2}b, cyan, yellow). The arrows N2, N4, and S2, S4, S5 indicate trajectories along which magnetic flux transformations occur. Indeed, domains of emergent magnetic flux appear near the base of the arrows. 
Stable CHs originate at mid- to high-latitudes (near the arrow ends).
The leading-polarity surges N3 and S3 disturb the regular magnetic flux transport. Small CHs appeared also within the leading-polarity surges. Thus, the CHs originated independently within the opposite polarity surges. This circumstance confirms a complicated nature of the polar fields reversal in Cycle~21. Generally, high-speed solar wind streams represent the final stage of magnetic flux evolution and its exit into the heliosphere.

\subsection{Spatiotemporal Evolution of the Sun's Magnetic Field in Cycles~21--24}
\label{subsec:time_evol}
\begin{figure*}
\centering
\includegraphics[scale=0.555]{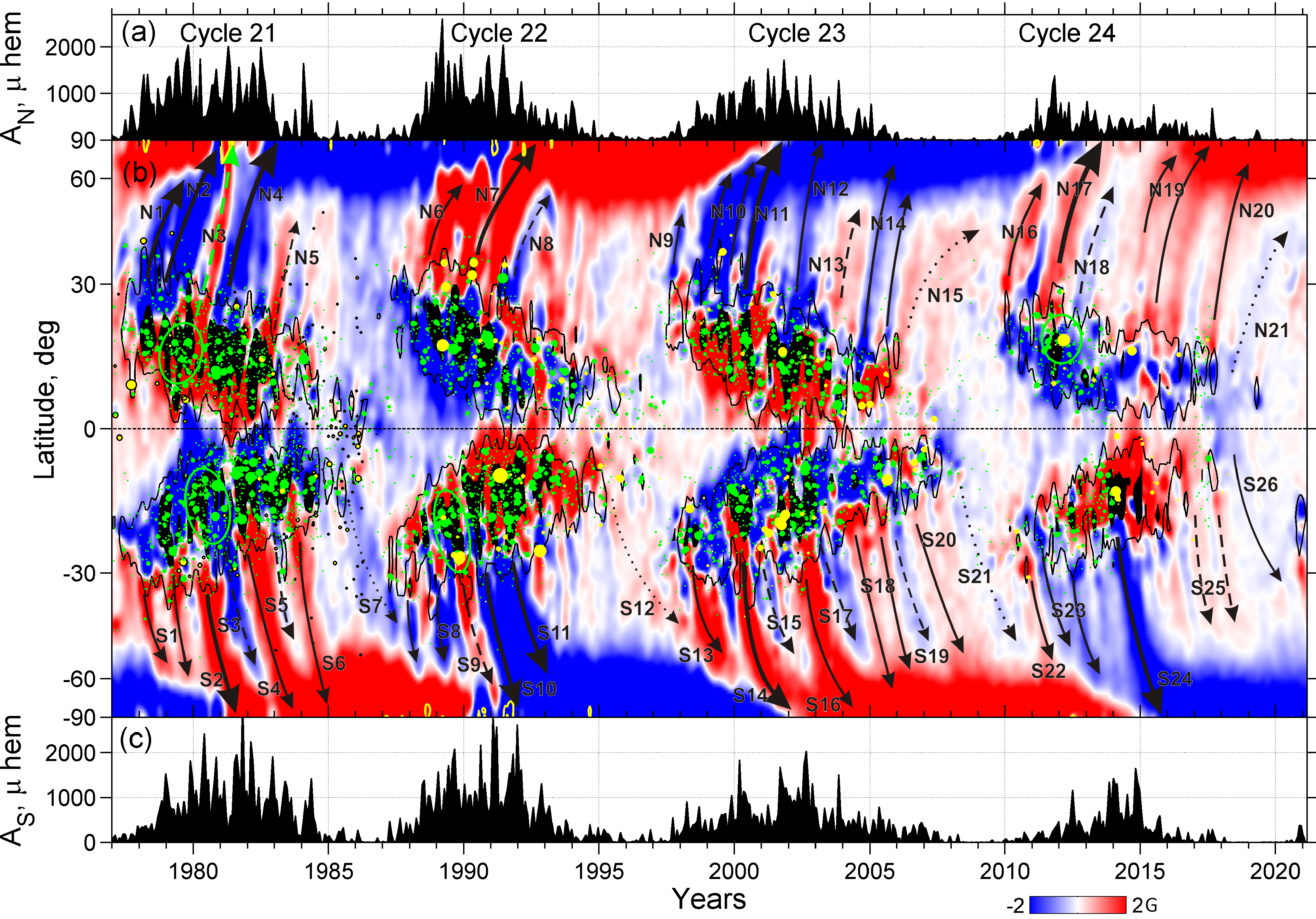}
\caption{
(a/c) Same as \Fig{figure2}. (b): Zonally averaged magnetic fields are shown in blue-to-red. Zones of intense sunspot activity are shown with black spots. Black contours depict the boundaries of sunspot activity. 
Solid/dashed arrows show the trailing/leading-polarity surges within the cycles,
while dotted arrows show the leading-polarity surges between adjacent cycles.
Non-Joy and anti-Hale ARs are marked with green/yellow markers whose diameter is proportional to sunspot areas. 
Yellow contours near poles show the `details' of the decomposition with magnetic flux density 2~G; see \Sec{subsec:time_evol} for details.
}
\label{figure3}
\end{figure*}

In this subsection, we present the time-latitude analysis of synoptic maps averaged over longitude for CRs 1625--2241 to study a spatiotemporal evolution of Sun's magnetic field in Cycles~21--24. 
We subtracted mean magnetic flux for every synoptic map.
The time-latitude distribution was denoised using wavelet-based technique
that recognizes domains of possible errors caused by the polar-field extrapolation.
We applied Multilevel 2-D Non-Decimated Wavelet Reconstruction with Haar wavelet and level of 4.
The main idea of this method is that the original time-latitude distribution 
is decomposed into `approximation' and `details' \citep{starck06}.
The approximation shows the evolution of large-scale magnetic fields (\Fig{figure3}b).
We have also examined the details which demonstrate the small-scale magnetic fields and possible defects of the time-latitude distribution due to the polar-field extrapolation. 
During 1977--1999, these details were concentrated in the AR areas and near the poles, more or less regularly.
In \Fig{figure3}b only the details with magnetic flux density of 2~G 
near the poles are displayed by yellow contours.
Near north pole, small-scale details usually occurred in the first half of every 
year. 
This is because the annually-varying angle between the solar rotation axis and our line of sight from (near) Earth 
causes the north/south pole to become unobservable during the first/second half of each year.

\subsubsection{Cycle~21}
Now, we consider Cycle~21 in a time-latitude aspect. 
After the decay of first ARs, the remnant flux surges N1/S1 were formed in the north/south 
(\Fig{figure3}b). These surges 
approached $\pm60\degr$ latitudes. Reconnection of opposite magnetic polarities 
led to their partial annihilation, the latitudinal extent of the polar UMRs decreased. 
In 1978--1979, large ACs occurred in both hemispheres (\Fig{figure3}a,c). During ARs evolution and decay, weak magnetic fields were dispersed in the surrounding photosphere, forming UMRs. After the decay of long-lived ACs, a surge of negative (trailing) polarity N2 was formed in the northern hemisphere. It is marked with a solid arrow. 
The poleward transport of the negative-polarity UMR 
changed the dominant polarity (+/$-$) in late 1980.
Subsequently,
a surge of positive (leading) polarity (N3, green arrow) 
produced at low-latitudes and moved to high latitudes 
in 1981. 
The low-latitude base of surge N3 is related to non-Joy ARs 
that were concentrated at latitudes 5--20$\degr$ in 1979. 
The event 
N3, however, occurred during the period of poor visibility of the north pole. 
Under such unusual conditions, the extrapolation led to significant errors, 
which manifested themselves in subpolar fields during 1981. 
Therefore, the north pole did not necessarily reverse around this time
and the pole filling technique might have imposed the sub-polar polarity reversals 
onto the polar fields in error.

\begin{figure*}
\centering
\includegraphics[scale=0.2]{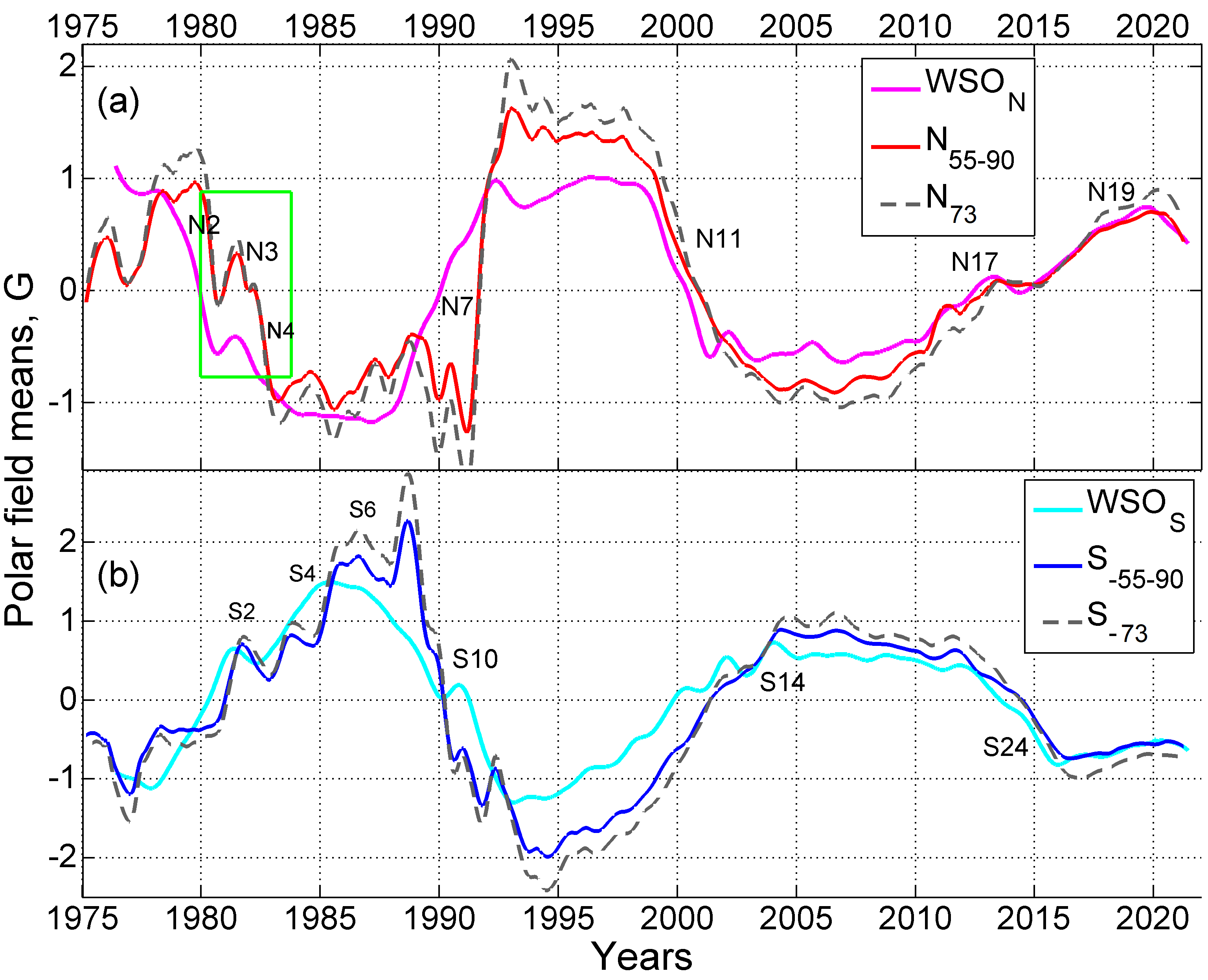}
\caption{
Magenta/cyan: north/south polar fields from the daily line-of-sight measurements 
with 20~nHz low pass filtered at
WSO. Red/blue: north/south polar-cap magnetic flux densities computed by
averaging of magnetic fields at latitudes since $\pm55^\circ$ to pole in the
radial synoptic maps. Gray dashed lines show the mean flux densities at
latitudes $\pm73^\circ$ in the maps. Here, for ease of comparison, the
synoptic map data are converted to the WSO scale using the conversion factors \citep{Riley2014}.
}
\label{figure4}
\end{figure*}

Large ARs were observed in 1980--1981. Their decay led to the
formation of a negative polarity (N4) surge, which reached the north
pole and 
strengthened the polar cap of negative polarity.
This complicated restructuring 
is also seen in the polar cap field of 
\Fig{figure4};
also see \citet{Janardhan18}.
The changes in polar fields originated due to the remnant flux surges 
N2, N3, N4 are shown in the green rectangle in \Fig{figure4}.
After the decay of anomalous ARs, UMRs of leading polarity were
formed at higher latitudes. Starting from higher latitudes, these
UMRs were transported poleward and resulted in the polar field
reversal. The analysis of the present high-resolution synoptic maps
essentially 
shows
similar features 
based on the WSO data as
reported in \citet{Mordvinov2019}.
The timings of the reversals of the dominant polarity are listed in \Tab{table1}.
As the data quality is not adequate to correctly determine the multiple reversals (if any) 
in the north pole for Cycle 21, we include the most likely one in \Tab{table1}.

In the southern hemisphere, a surge of the trailing (positive)
polarity formed at the beginning of the cycle. The critical surge
(S2) reached the south pole and led to a change in the dominant
polarity ($-$/+). Later, a surge of the leading (negative) polarity
(S3) was formed. It approached the south pole, but there was no
change in the dominant polarity (also see \Fig{figure4}). A few
abnormal ARs were observed at the base of this surge.

During 1982--1984, the decay of ARs led to the formation of intense
surges S4 and S6, which restored and strengthened the magnetic field
of positive polarity at the south pole. The latitudinal extent of
the polar cap increased. There were many anomalous ARs in Cycle~21
\citep{WS89}. Their decay led to the formation of leading-polarity
surges N3, N5, S3, and S5. The high level of magnetic activity and
much of anomalous ARs resulted in 
the complicated restructuring of the subpolar flux.

The surge S7 is of particular interest. It originated from
leading-polarity UMR at the end of Cycle~21. Its path indicates the
magnetic flux transport from Cycle~21 and merged with new magnetic
flux of the same polarity in the next cycle. This merger leads to
an increase in the magnetic field in a new cycle and early reversal
in south (also see \Fig{figure4}). This phenomenon indicates that not all
of the magnetic flux disappears by the end of the old cycle; part
of it can be transferred from one cycle to the next and amplify the
field.

\subsubsection{Cycle~22}
In Cycle~22, 
after the decay of the first ARs, trailing-polarity surges N6 and S8 were formed (\Fig{figure3}b). 
These surges annihilated at about $\pm60^\circ$ latitudes, 
and thereby reduced the polar magnetic flux. During 1989--1992, 
plenty of sunspots appeared. 
The critical surges N7 and S10 
reached the poles and changed the dominant polarity. 
In the southern hemisphere, there were intense surges S10, S11 which
represent cumulative effects of two activity impulses (\Fig{figure3}c).
In both hemispheres, regular polar field reversals occurred. Such
an evolution of magnetic field is quite consistent with the
Babcock--Leighton scenario.

Nevertheless, these patterns were disrupted by leading-polarity 
surges. In both hemispheres well-defined surges N8 and S9 of leading polarity occurred. 
At the base of these surges, abnormal ARs were observed. 
The leading-polarity surge S9 was not strong enough to cause
a clear dip
in the polar-cap field in \Fig{figure4} around 1991.
Afterward around 1992 in \Fig{figure4}, there was a clear depression in the south polar-cap field, 
the signature of which is not visible in the time-latitude 
plot.
This depression in the polar-cap field caused a double peak in the 
following sunspot Cycle~23 of the southern hemisphere. 
This connection between the polar field depression and the double peak 
was demonstrated in \citet{KMB18} based on the observations and dynamo model. 
As there is no prominent drop in the polar field in the northern hemisphere, 
no double peak seen in the same hemisphere for Cycle~23.

Again, in the southern hemisphere of Cycle~22, we 
find 
a link between adjacent cycles. After the decay of the last ARs, 
a surge of (positive) leading-polarity S12 was formed
and transported in mid-latitudes until it merged with S13, 
which had the same polarity as that in Cycle~23.

\subsubsection{Cycle~23}
In 1997, the decay of first ARs appeared around $\pm30\degr$ 
latitudes form UMRs of both polarities (\Fig{figure3}b). Trailing polarity UMR N9 is well-defined.
The decay of subsequent ARs was associated with surges N10 that reached higher latitudes. 
The decay of long-lived ARs in 2000 led to the formation of the critical surge N11, which reached the north pole and led to a change in the dominant polarity (+/$-$). Subsequent bursts of activity were associated with 
surges N12, N14, that strengthened magnetic flux at the north pole. 
The leading-polarity surge N13 
is apparently associated with non-Joy ARs at low latitudes.

At the end of Cycle~23, the leading-polarity UMRs dominated near
the equator. 
During the activity minimum, the
meridional flow led to formation of an extended surge of the remnant
flux N15, which reached latitudes of $40-60\degr$ by the beginning
of Cycle~24. Subsequently, a surge N16 of a new cycle was formed
at 
low
latitudes. The merger of the positive polarity UMRs led
to an increase in the net remnant flux of the new cycle. Thus, again
the transport of magnetic flux between adjacent cycles reveals a
new mechanism that links individual cycles into pairs.

In south, a complex structure of remnant flux was observed. 
In 2000, decay of great ARs generates critical surge S14. 
Subsequent trailing-polarity surges S16, S18, and S20 strengthened the magnetic flux in the polar region.
The leading-polarity surges S15, S17, and S19 tried to reduce the polar flux.
Surge S17 clearly caused a little depression in the polar-cap field as seen in \Fig{figure4}. 
At the end of Cycle~23, leading-polarity surge S21 was formed. 

We finally notice that as there is no prominent depression in the
north/south polar field in Cycle~23 (\Fig{figure4}), there are
no pronounced double peaks in the following cycle.

\subsubsection{Cycle~24}
This is the most important cycle because the magnetic activity was
low, and this field is the precursor for the upcoming Cycle~25. The
north-south asymmetry of the Sun's activity has led to a significant
asynchronous reversal of the polar fields. At the beginning of the
cycle, magnetic activity prevailed in the northern hemisphere.
During the decay of the first ARs in the northern hemisphere, surge
N16 was formed (\Fig{figure3}b). The critical surge N17 reached the north pole and
changed the dominant polarity ($-$/+) in early 2013.

During 2012--2013, the activity was low and a leading-polarity surge
N18 was formed.  During that time, anomalous ARs were observed at
low latitudes.  As surge N18 approached the north
pole, magnetic flux decreased to almost zero due to annihilation
of opposite polarities (also see \Fig{figure4}). The decay
of ARs in 2014--2017 resulted in surges N19, N20. The poleward
transport of remnant flux led to significant strengthening of the
polar field in the north pole. At the end of the cycle, the
leading-polarity surge N21 was formed. This surge reached mid-latitudes
by 2021. It already corresponds to upcoming Cycle~25.

At the beginning of cycle in southern hemisphere, 
decay of anti-Hale and non-Joy ARs led to the formation of a leading-polarity surge S22.
Its poleward transport overpowers the previous polarity polar field and destroys the weak surge S21 
from the previous cycle. Further appearance of ARs and their decay led to the formation of trailing-polarity surges S23. The emergent magnetic flux peaked in 2014 for south. 
After the decay of largest ARs in 2014, a critical surge S24 was formed. 
This surge reached the south pole and changed the dominant polarity by mid-2015 (+/$-$). 
During the decline phase of activity, leading-polarity surges S25 were formed. 
Finally, surge S26 originated from the transequatorial UMR in the southern
hemisphere.

\begin{table}
\caption{
Changes in the dominant magnetic fields in the subpolar zones for Cycles~21--24.
}
\begin{center}
\begin{tabular}{lcccc}
\hline
\hline
Cycle number & 21                     & 22      & 23     & 24 \\
\hline
North pole & 1982.5 & 1991.7 & 2001.8 & 2013.3 \\
\hline
South pole & 1981.1 & 1990.2 & 2001.9 & 2015.4 \\
\hline
\end{tabular}
\end{center}
\label{table1}
\end{table}


\begin{figure*}
\centering
\includegraphics[scale=0.33]{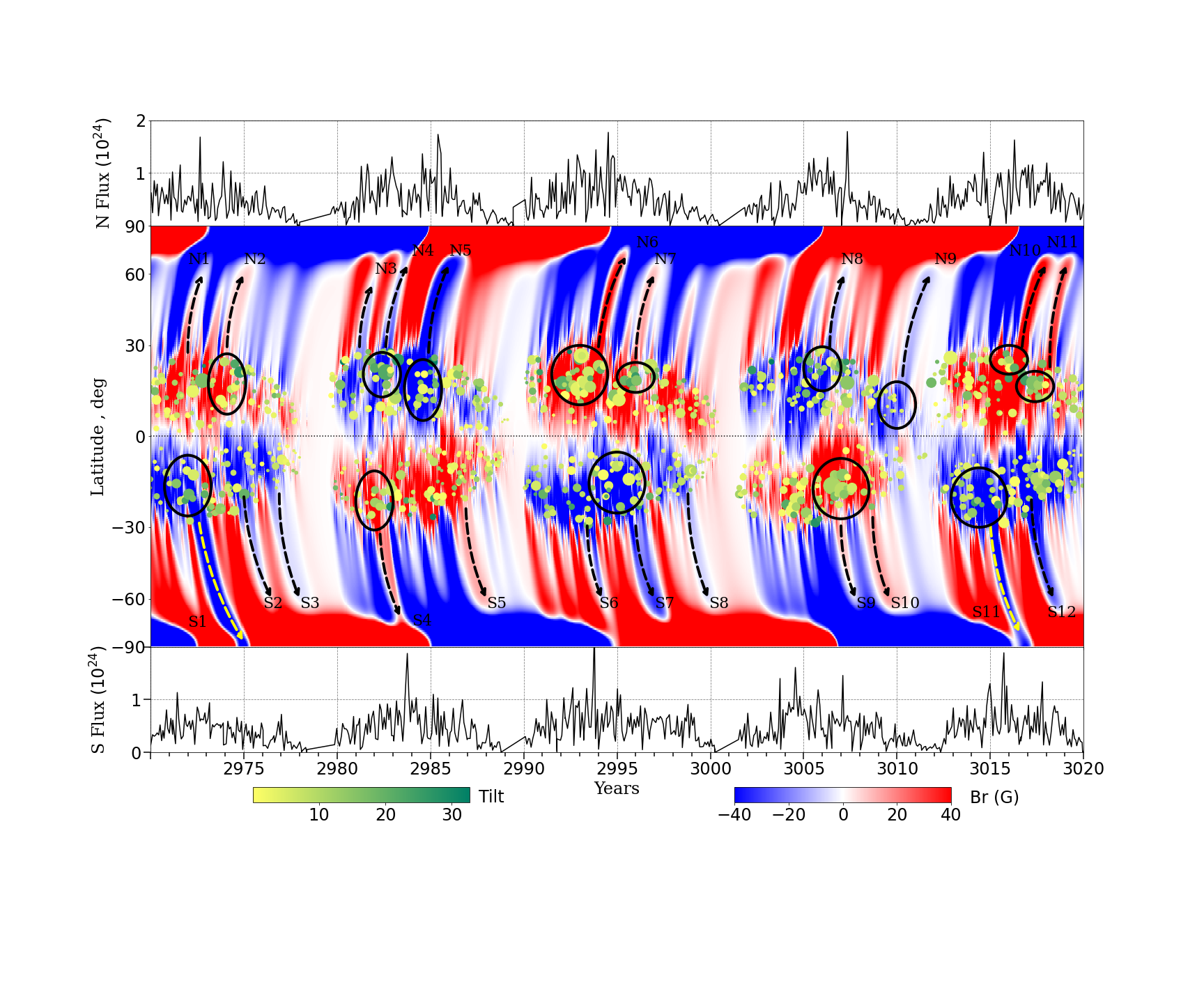}
\caption{
Results from dynamo model:
The butterfly diagram of the radial magnetic field in Gauss (middle).
Filled points represent anti-Hale BMRs with tilt deviating from Joy's law by more than $50\%$. 
The colorbar shows the modulus of the tilt.
The size of the points indicates the amount of flux, while color shows 
the amount of absolute tilt (in degree); see left colorbar.
Top and bottom panels show the variations of the monthly magnetic fluxes of BMRs in the northern and 
southern hemispheres in Mx.
}
\label{fig:mod}
\end{figure*}

\section{Leading polarity surges and a triple reversal in a Babcock--Leighton dynamo model}
Here we present a snapshot of the surface radial magnetic field from a dynamo model to explore and compare
the results with the observational ones.
Our result is obtained from Run~B1 which was presented in \citet{Kar20}
and it is based on the 3D dynamo model, STABLE \citep{MD14,MT16,KM17}.
The details of the model which produces our result are described in \citet{Kar20}.
Therefore, without writing the mathematical details of the model, we
mention here the salient features. Essentially, in this model,
we solve the kinematic dynamo equation by specifying single-cell meridional flow,
differential rotation, turbulent diffusivity, and magnetic pumping, all consistent with
the available solar observations \citep{KC16}. The model includes a sophisticated procedure
for the BMR eruptions on the surface, based on the toroidal field at the tachocline.
The eruption is possible only when the time delay between the eruptions 
exceeds a certain delay $\Delta$
and the 
mean azimuthal field in the tachocline exceeds a certain threshold value $B_t$.
The $\Delta$ is obtained from a log-normal distribution which is fixed in this simulation. 
While $B_t = 2$~kG at the equator, it increases exponentially with the latitude. This helps to restrict BMRs in low- to mid-latitudes and importantly, the observed 
cycle-dependent latitude distribution of BMR, namely the stronger cycle (on average) 
produces BMRs at higher latitudes
than the weaker ones. The latter feature is sufficient to produce a 
nonlinearity to halt the growth
of the dynamo \citep{Kar20}. 
Statistical features of BMRs are obtained from observations, 
particularly the tilt is obtained from Joy's law with a Gaussian scatter around it having $\mu=0$ and $\sigma = 15\degr$ \citep{SK12, Jha20}.

In \Fig{fig:mod}, we present the surface radial field as a function of time and latitude.
One discrepancy in this model is that the strength of the radial field is much stronger than that is observed in the Sun.
Despite this, we observe the basic features of the solar cycle, including, polarity reversal, pole-ward
migration and hemispheric asymmetry. We also observe a lot of details of the magnetic field at small scales which are in
good agreement with the observation. The leading polarity surges which are originated at low- to mid-latitudes, migrate, and merge with the global field near the pole. Some of these leading polarity surges are so strong that they disturb the polar field significantly. \citet{KM17, HCM17, KM18} showed that the anti-Hale spots
are responsible for changing the polar field and consequently the next cycle strength in this dynamo model. 
The anti-Hale BMRs having tilt deviation more than $50\%$ from Joy's law are shown by color points in \Fig{fig:mod}.
We observe some connection (although it is not always obvious in this plot) between the
domains of highly tilted BMRs with large flux contents and the leading polarity surges,
as marked by dashed arrows.
As the time delay between two BMRs is not fixed and it is taken from an observed distribution,
the rate of BMR eruption is not constant.
This leads to some variation in the magnetic field.
In this figure, we find clear evidence of a triple reversal
in the southern polar field around the year 2975,
which is caused by a strong leading polarity surge S1 (yellow dashed arrow) 
that originated around $15\degr$ latitude.
Therefore, although the triple reversal in the Sun is not conclusively known 
due to the defects in the data of the polar regions,
it may be possible in the Sun as supported by a Babcock--Leighton dynamo model.
The surge S6 that appeared around the year 3016 in the southern hemisphere is very interesting.
It enhanced the old polarity (negative) field and caused a delay in the reversal
of the pole.
Had this surge appeared after a few months, this could lead to another triple reversal of the south pole.
We note that the triple reversal in the dynamo model is very rare.
In our simulation of around 500~years, we find one clear triple reversal
which is shown in \Fig{fig:mod}. In future work, we shall present a detailed analysis of
the statics of triple reveals in this dynamo model, by running the model for several thousands of years in different parameters.

\section{Summary and Conclusions}
We analyzed the high-resolution synoptic magnetic maps
for Cycles~21--24. We identify major remnant flux
surges and their sources, reexamine the polar field buildup and
reversals in their causal relation to the Sun's low-latitude activity.
A spatiotemporal behavior of the Sun's magnetic flux depicts a
general evolution of the Sun's magnetic field and its effect on
PCHs.


The time-latitude analysis of AR tilts showed that surges of leading
polarity appeared after the decay of non-Joy and anti-Hale ARs.
When such an AR is highly tilted and appears near the equator, it produces
a strong leading polarity surge which can disturb a dominant 
polarity flux considerably \citep{Ca13, KM17, Nagy17}.
We also found inter-cyclic remnant flux surges between
adjacent 11-year cycles. These surges reveal the physical links
between subsequent solar cycles. The individual 11-year cycles are
linked in pairs. This interrelation between 11-year cycles suggests
a possible long-term memory in solar activity that manifests itself
on a secular timescale.

Preliminary analyses of the magnetic field from a 3D Babcock--Leighton
dynamo model demonstrate some support for the generation of leading-polarity surges
at low-latitudes, their transport towards poles and changes in the dominant polarity field.
The model also shows a clear evidence of triple reversal, although it is very rare.
Therefore, although the triple reversal in the Sun is not very conclusive due to the known defects in the data of the polar regions, such phenomenon is not fundamentally impossible from the point of view of the Babcock--Leighton dynamo model.

\section*{Acknowledgment}
We thank Luca Bertello for the discussion on the limitations in the polar region data and the anonymous referee for raising some critical comments which helped to clarify
some points in the manuscript.
The work was financially supported by the Ministry of Science and Higher Education of the Russian Federation (M.A.V., G.E.M, Kh.A.I., Zh.A.V), RFBR (Grant No. 19-52-45002 M.A.V., G.E.M, Kh.A.I.), and Indo-Russian Joint Research Program of Department of Science and Technology with project number INT/RUS/RFBR/383 (Indian side). 

\section*{DATA AVAILABILITY}
NSO/Kitt Peak data used here are produced cooperatively by NSF/NOAO, NASA/GSFC, NOAA/SEL, and Bartol. Data were acquired by SOLIS instruments operated by NISP/NSO/AURA/NSF. SOHO is a project of international cooperation between ESA and NASA. The SDO/HMI data are available by courtesy of NASA/SDO and the AIA, EVE, and HMI science teams. 
"Bipolar magnetic regions determined from KPVT full disk magnetograms" were downloaded from the solar dynamo dataverse (https://dataverse.harvard.edu/dataverse/solardynamo), maintained by Andrés Mu\~noz-Jaramillo.
Data and the routines of our analyses presented in the article will be shared upon reasonable request to the corresponding author.
Wilcox Solar Observatory data used in this study was obtained via the web site http://wso.stanford.edu courtesy of J.T. Hoeksema. The Wilcox Solar Observatory is currently supported by NASA.

\bibliographystyle{mnras}
\bibliography{paper}

\end{document}